\documentclass[12pt]{article}

\usepackage{epsfig}

\newcommand{\sect}[1]{\setcounter{equation}{0}\section{#1}}

\def\be{\begin{equation}}
\def\ee{\end{equation}}
\def\ba{\begin{eqnarray}}
\def\ea{\end{eqnarray}}

\title{{\bf Dynamic Dilatonic Domain Walls}}
\author{H.A. Chamblin\thanks{email: H.A. Chamblin@damtp.cam.ac.uk} and
H.S. Reall\thanks{email: H.S.Reall@damtp.cam.ac.uk}
 \\ DAMTP \\ University of Cambridge \\ Silver Street,
Cambridge CB3 9EW, United Kingdom. \\ \\ Preprint DAMTP-1999-35}

\date{March 25, 1999}

\begin{document}

\maketitle

\begin{abstract}

Motivated by the ``universe as a brane'' idea, we investigate the
motion of a $(D-2)$-brane (or domain wall) that couples to bulk
matter. Usually one would expect the spacetime outside
such a wall to be time dependent however we show that in certain cases
it can be static, with consistency of the Israel equations yielding
relationships between the bulk metric and matter that can be used as
ans\"atze to solve the Einstein equations. As a concrete model we
study a domain wall coupled to a bulk dilaton with Liouville
potentials for the dilaton both in the bulk and on the wall. The bulk
solutions we find are all singular. Some have black hole or
cosmological horizons, beyond which our solutions describe domain
walls moving in time dependent bulks. A significant period of world
volume inflation occurs if the potential on the wall is not too steep;
in some cases the bulk also inflates (with the wall comoving) while in
others the wall moves relative to a non-inflating bulk. We apply our
method to obtain cosmological solutions of Ho\v{r}ava-Witten theory
compactified on a Calabi-Yau space. 

\end{abstract}

\sect{Introduction}

A domain wall in a $D$-dimensional spacetime
is an extended object with $D-2$ spatial dimensions which partitions
the spacetime into different domains. 
In cosmology, the different domains might
correspond to different vacua of some Higgs field. We will use the
term ``domain wall'' loosely to refer to any $(D-2)$-brane moving in 
$D$ dimensions.

When the gravitational back-reaction of a domain wall is included,
the global causal structure of the resulting spacetime is usually
modified.  In the simplest models, one considers domain walls which
do not couple to any bulk fields with the domain wall world volume
governed by a Nambu-Goto action
\be
 S_{DW} = -{\mu}{\int}_{\Sigma}d^{D-1}x\sqrt{-h}
\ee
where $\mu$ is a parameter which corresponds to the tension or
energy density of the wall, $\Sigma$ denotes the world volume swept out by the
wall and $h$ denotes the determinant of the world volume metric. 
More precisely, the stress-energy tensor
$T_{MN}$ of the domain wall takes the simple form
\be
 T_{MN}=\mu\delta(x)\textrm{diag}(1,-1,\ldots,-1,0),
\ee
where $x$ is the direction transverse
to the wall.  This stress-energy tensor is a distributional source
for the Einstein equations: the metric is at most $C^0$ as we move across
such a ``thin wall'' of energy density.  One approach to understanding the
gravitational field of such an object is to divide the spacetime into smooth
domains, where each domain is bounded by a domain wall. If we vary the
metric to obtain the equations of motion then we obtain the well-known
{\it Israel matching conditions} \cite{israel}:
\be
\label{eqn:is}
 \{K_{MN}-Kh_{MN}\}=8\pi G\mu h_{MN},
\ee
where $h_{MN}$ is the induced metric on the domain wall, $K_{MN}$ its
extrinsic curvature and $K=h^{MN}K_{MN}$. The extrinsic curvature is
typically discontinuous across a domain wall; the curly brackets
denote summation over each side of the wall\footnote{Our convention is
that the normal to domain wall points into the bulk on \emph{both}
sides, so we take a sum over the sides of the wall rather than a difference.}.
We present a careful derivation of the Israel conditions from the
Einstein-Hilbert action in section \ref{sec:derive}.

The Israel conditions are often used to study the motion of a domain
wall in a static bulk spacetime. They are satisfied by seeking totally
umbilic ($K_{MN} \propto h_{MN}$) surfaces in the bulk, slicing along
such a surface and then gluing to another such bulk. For example,
domain walls between domains corresponding to different vacua of a
Higgs field can be obtained by taking the bulk spacetime to be flat
\cite{ber,vil,ip} or 
false vacuum decay can be studied by taking the two bulk portions to
have differing cosmological constants \cite{ber,guth}. One can have a
domain wall enclosing a bubble of true vacuum in a sea of false
vacuum. The Israel conditions reduce to two equations, one giving the
velocity of the domain wall and the other its
acceleration. Consistency of these equations follows from the bulk
Einstein equations.

If the domain wall action is of the Nambu-Goto form considered above
then the energy density on the wall is fixed so there can be no
transfer of energy between the bulk matter and the wall. However it is
possible to consider more general situations in which matter is
localised to the wall and energy can flow on or off the wall. In this
case, the right hand side of equation \ref{eqn:is} is replaced by the
energy-momentum tensor of the wall matter. For simple static bulk
solutions the Israel conditions reduce to two equations, one relating
the velocity of the wall to its energy density and the other relating
its acceleration to its energy density and pressure. There is also an
equation of state for the matter on the wall. Thus there are three
equations for three quantities (position, energy density and pressure)
so one would expect a solution to exist without further restrictions
on the bulk spacetime.

This changes if the domain wall couples to matter in the bulk. For
example, in string theory a brane or domain wall will usually be
coupled to a dilaton $\phi$ - a bulk scalar field which 
measures the 
scale or deformation properties of some internal manifold.
The simple Nambu-Goto form for the world volume action is replaced
by an action of the form
\be
 S_{DW} = -{\int}_{\Sigma}d^{D-1}x\sqrt{-h}\hat{V}(\phi),
\ee
where the wall tension $\hat{V}$ depends on the value of the
dilaton on the wall. If the domain wall moves through regions of
varying dilaton then energy will flow on or off the wall. If one
attempts to find solutions for the domain wall motion as before then
one encounters a problem. Once again there are three equations for three
quantities but now the energy density and pressure on the wall are
specified by its position in the bulk. Hence there is no guarantee
that a solution will exist for the motion of the wall in a static bulk
spacetime: in general the motion of the domain wall will make the bulk
time dependent. Studying dynamics in this model is hard because the
bulk metric will depend on two parameters (time and distance from the
wall) so the Einstein equations become more difficult to solve.

In this paper we investigate the circumstances under which it \emph{is}
possible for a domain wall coupled to bulk matter to move in a static
bulk. Starting from a static ansatz for the bulk metric we find in
section \ref{sec:israel} that 
consistency of the Israel equations requires that
the bulk metric and matter be related in a certain way. This can be
used to find solutions of the bulk Einstein equations. We concentrate
on the example of a bulk dilaton with Liouville potentials in the bulk
and on the domain wall. This model can be motivated by massive
supergravity theories and p-brane world volume actions.
Domain walls in such models have been extensively studied for static
bulk spacetimes \cite{cvetic1,cvetic2} (see \cite{cvetic3} for a
review), especially in the supersymmetric case
\cite{cvetic1,lu}. However all of these solutions assume a constant
dilaton on the domain wall whereas in our solutions the dilaton
evolves in time on the wall. 

Our bulk solutions are described in section \ref{sec:bulk}. All are
singular but in some cases the singularity is hidden behind a horizon.
Some have cosmological horizons beyond which the
metric becomes time dependent. The motion of the domain wall can be
followed across the horizon so some of our solutions describe the
evolution of a  domain wall in a cosmological background. 

\bigskip

There has been recent interest in cosmological models in which our universe is
viewed as a brane moving in a higher dimensional spacetime, possibly
with a very low fundamental Planck scale and consequently large extra
dimensions \cite{large1}. A topic of particular interest in this
scenario is how inflation occurs on the brane. In \cite{linde} it was
concluded that for a viable model the bulk must be non-static
during inflation. However in this model inflation was assumed to be
driven by a scalar field restricted to the brane world volume and the
possibility of the energy density on the brane coming from a bulk
field was not discussed. In \cite{dvali} it was
described how inflation can occur if one of a stack of branes is
displaced from the others. In the case of a stack of
D-branes, the energy density that drives world volume inflation arises
from the energy of open strings stretched between the branes. This
energy can be viewed as coming from the non-zero expectation value of 
a world volume Higgs field. 
Inflation was discussed in \cite{large2}
where the bulk spacetime was assumed to be non-static but the
gravitational back reaction of the domain wall was treated in an
approximate manner that didn't correspond to a wall localized in the
extra dimensions. In \cite{lukas3} solutions were given for localized
domain walls with inflation driven by matter living on the wall. 
Our model can be used to study inflation driven by energy density on
the domain wall coming from a bulk field, taking full account of
the gravitational back reaction of the wall.

We find in section \ref{sec:results} 
that power law inflation can occur on the domain wall provided
that the potential on the world volume is not too steep. 
In some cases inflation occurs because the domain wall is comoving with an
inflationary bulk while in others it occurs because the domain wall
moves relative to the bulk. 

\bigskip

In section \ref{sec:higherdim} we show that 
for some values of the parameters our model can be obtained by
dimensional reduction of a Nambu-Goto domain wall moving in a bulk
spacetime with a cosmological constant. The world volumes of these
domain walls undergo exponential inflation, which gives the power law
inflation described above in the dimensionally reduced
theory.

Cosmological solutions of Ho\v{r}ava-Witten theory, the strongly
coupled limit of the $E_8 \times E_8$ heterotic superstring theory,
have been recently discussed in \cite{benakli,lukas2,hsr}.
The orbifold fixed planes in this theory can be viewed as domain
walls so our method is well suited to finding new solutions. We discuss
these in section \ref{sec:hw}.

In section \ref{sec:conclude} we discuss our conclusions and speculate
on possible generalisations and applications of our work to cosmology
and the dynamics of branes in string theory. 

\sect{The Israel Matching Conditions}

\label{sec:derive}

This section consists of a simple derivation of the Israel conditions
\cite{israel} for matching the metric across a domain wall. The reader
familiar with this derivation is advised to skip to the next section.

Let $M$ be
a $D$ dimensional manifold containing a domain wall $\Sigma$, which
splits $M$ into two parts, $M_{\pm{}}$. We demand that the metric be
continuous everywhere and that the derivatives of the metric be
continuous everywhere except on $\Sigma$. We shall denote the two
sides of $\Sigma$ as $\Sigma_{\pm{}}$.

Varying the Einstein-Hilbert action in $M_{\pm{}}$ gives\footnote{We
use units in which $8\pi G=1$, a positive signature metric and a
curvature convention for which de Sitter space has a positive Ricci scalar.}
\be
 \delta S_{EH}=-\frac{1}{2} \int_{\Sigma_{\pm{}}} d^{D-1}x \sqrt{-h}
g^{MN}n^P(\nabla _M \delta g_{NP} - \nabla_P \delta g_{MN}),
\ee
where $n_M$ is the unit normal pointing \emph{into} $M_{\pm{}}$ and the
induced metric on $\Sigma_{\pm{}}$ is given by the tangential components
of the projection tensor $h_{MN}=g_{MN}-n_M n_N$. Note that the
quantity in brackets vanishes when contracted with $n^Mn^Nn^P$ so we
can replace $g^{MN}$ by $h^{MN}$ to get
\be
 \delta S_{EH}=-\frac{1}{2} \int_{\partial M} d^{D-1}x \sqrt{-h}
h^{MN}n^P(\nabla _M \delta g_{NP} - \nabla_P \delta g_{MN}).
\ee
This expression contains a normal derivative of the metric variation,
which we are allowing to be discontinuous across
$\Sigma$, so the contributions from the two bulk regions will not
necessarily cancel. Therefore it is necessary to include a
Gibbons-Hawking boundary term \cite{gibbons} to cancel this term.
On each side of the domain wall we include a contribution to the
action
\be
 S_{GH}=-\int_{\Sigma_{\pm{}}}d^{D-1}x \sqrt{-h} K,
\ee
where $K$ is the trace of the extrinsic curvature of $\Sigma_{\pm{}}$
i.e. $K=h^{MN}K_{MN}$, $K_{MN}=h_M^P h_N^Q \nabla_P n_Q$. 

We now have to compute the variation of this new term. This is
slightly complicated by the fact that $n_M$ depends on the metric
because it is normalised to unit length. When the metric is varied,
the change in $n_M$ is
\be
 \delta n_M=\frac{1}{2} n_M n^P n^Q \delta g_{PQ}.
\ee
The variation in $K$ is
\be
 \delta K=-K^{MN}\delta g_{MN} - h^{MN}n^P(\nabla_M \delta g_{NP} -
\frac{1}{2}\nabla_P \delta g_{MN}) + \frac{1}{2}K n^Pn^Q\delta g_{PQ},
\ee
and the variation of the Gibbons-Hawking term is
\be
 \delta S_{GH}=-\int_{\Sigma_{\pm{}}}d^{D-1}x \sqrt{-h} (\delta K +
\frac{1}{2}K h^{MN}\delta g_{MN}).
\ee
Note that $\delta K$ contains a term
$\frac{1}{2}h^{MN}n^P\nabla_P\delta g_{MN}$, which cancels the
corresponding normal derivative in the variation of the 
Einstein-Hilbert action. The total variation is
\ba
\label{eqn:deltaS}
 \delta S_{EH}+\delta S_{GH} & = & \int_{\Sigma_{\pm{}}} d^{D-1}x\sqrt{-h}
\left[\frac{1}{2}h^{MN}n^P\nabla_M\delta g_{NP} + K^{MN}\delta g_{MN}
- {} \right. \nonumber \\ {} & - &\left. \frac{1}{2}Kn^Mn^N\delta g_{MN}
-\frac{1}{2}K h^{MN}\delta g_{MN}\right].
\ea

To proceed further we need the following simple result for a vector
field $X^M$ tangential to $\Sigma_{\pm{}}$
\be
 \nabla_M X^M = h^{MN}\nabla_M X_N + n^Mn^N\nabla_M X_N =
\bar{\nabla}_M X^M - X^M n^N \nabla_N X_M,
\ee
where $\bar{\nabla}$ is the covariant derivative associated with the
induced metric on $\Sigma_{\pm{}}$. Using this and the definition of
$K_{MN}$ gives
\ba
 h^{MN} n^P \nabla_M \delta g_{NP}  & = & \nabla_M(h^{MN}n^P\delta
g_{NP})-\delta g_{NP}\nabla_M(h^{MN}n^P) \\ \nonumber & = &
\bar{\nabla}_M (h^{MN}n^P\delta g_{NP}) + Kn^Mn^N\delta g_{MN}
-K^{MN}\delta g_{MN}.
\ea
Finally this can be substituted into $\ref{eqn:deltaS}$ and the total
derivative term integrated away to give
\be
 \delta S_{EH}+\delta S_{GH}=\frac{1}{2}\int_{\Sigma_{\pm{}}}
d^{D-1}x\sqrt{-h}\left(K^{MN}-K h^{MN}\right)\delta g_{MN}.
\ee
Note that $K_{MN}$ can be discontinuous across $\Sigma$ so the
contributions from $\Sigma_{\pm{}}$ need not cancel.

If the domain wall has action
\be
 S_{DW}=\int_{\Sigma}d^{D-1}x\sqrt{-h}L_{DW}
\ee
then its variation is
\be
 \delta S_{DW}=\int_{\Sigma}d^{D-1}x\sqrt{-h} t^{MN}\delta g_{MN},
\ee
where we have used the fact that $t^{MN} \equiv
\frac{2}{\sqrt{-h}}\frac{\delta S_{DW}}{\delta h_{MN}}$ is tangential to
the domain wall. The variation of the total action
$S=S_{EH}+S_{GH}+S_{DW}$ gives the Israel conditions
\be
 \left\{K_{MN}-K h_{MN}\right\}=-t_{MN},
\ee
where the curly brackets denote summation over both sides of $\Sigma$.

The stress energy tensor $t^{MN}$ on the domain wall is not
necessarily conserved because energy can flow between the domain wall
and the bulk. This can be seen by taking the divergence of the Israel
equations:
\be
 \bar{\nabla}_M t^{MN}=-\left\{\bar{\nabla}_M
K^{MN}-h^{MN}\bar{\nabla}_M K\right\}.
\ee
The right hand side can be evaluated using Codacci's equation
\cite{hawking}, giving
\ba
\label{eqn:cons}
 \bar{\nabla}_M t^{MN} & = &-\left\{h^{NM}R_{MP}n^P\right\} \nonumber
\\ & = &-\left\{h^{NM}T_{MP}n^P\right\},
\ea 
where $T^{MN}$ is the bulk energy momentum tensor and we have made use
of the bulk Einstein equation. This equation describes conservation of
energy when it moves from the bulk to the boundary or vice versa.

\sect{The Equations of Motion}

\label{sec:model}

The example that we shall be focussing on is Einstein gravity
with a scalar field (dilaton) in the bulk and a domain wall that couples
to the bulk dilaton. The action is
\be
 S=\int_M
d^{D}x\sqrt{-g}\left(\frac{1}{2}R - \frac{1}{2}(\partial\phi)^2 -
V(\phi)\right) + S_{DW},
\ee
where
\be
 S_{DW}=-\int_{\Sigma}d^{D-1}x\sqrt{-h}(\{K\}+\hat{V}(\phi)).
\ee
Note that we have absorbed the Gibbons-Hawking terms into the domain
wall action. Once again, curly brackets denote summation over both
sides of the wall.
The bulk Einstein equation is
\be
 R_{MN}=\partial_M\phi\partial_N\phi +\frac{2}{D-2}V(\phi)g_{MN}.
\ee
Varying the scalar field gives
\be
 \int_M d^D{x}\sqrt{-g}\left(\nabla^2\phi -
\frac{dV}{d\phi}\right)\delta\phi +
\int_{\Sigma}d^{D-1}x\sqrt{-h}\left(\{n.\partial\phi\}
-\frac{d\hat{V}}{d\phi}\right)\delta\phi = 0
\ee
because $n$ points \emph{into} the bulk. This yields the equation of motion
\be
 \nabla^2\phi=\frac{dV}{d\phi}
\ee
with a boundary condition at the domain wall
\be
 \label{eqn:dilbc}
 \{n.\partial\phi\}=\frac{d\hat{V}}{d\phi}.
\ee
This ensures that the energy conservation equation \ref{eqn:cons} is
satisfied.

We also need the Israel equations, which can be written
\be
 \{K_{MN}\}=-\frac{1}{D-2}\hat{V}(\phi)h_{MN}.
\ee

\sect{Domain Wall Motion in a Static Background}

\label{sec:israel}

 We shall seek solutions in which the bulk
spacetime is symmetric under reflection in the domain wall, hence
the extrinsic curvatures on each side of the wall are the same and the
Israel equations become
\be
 K_{MN}=-\frac{1}{2(D-2)}\hat{V}(\phi)h_{MN}.
\ee 
In general there is no reason to assume that a moving domain wall will
give rise to a static bulk spacetime. However it is sometimes possible
that this can occur. Consider a bulk metric
\be
\label{eqn:metric}
 ds^2=-A(r)dt^2+B(r)dr^2+R(r)^2d\Omega_k^2,
\ee
where $d\Omega_k^2$ is the line element on a $D-2$ dimensional space
of constant curvature with metric $\bar{g}_{mn}$ and Ricci tensor
$\bar{R}_{mn}=k(D-3)\bar{g}_{mn}$ with $k \in \{-1,0,1\}$.
We shall also include a dilaton $\phi(r)$.

Let the position of the domain wall be $r=r(t)$ with the above metric
valid on the $r<r(t)$ parts of surfaces of constant $t$ 
and its reflection valid on the $r>r(t)$ parts. The unit
normal pointing into $r<r(t)$ is
\be
 n_M=\frac{\sqrt{AB}}{\sqrt{A-B\dot{r}^2}}(\dot{r},-1,0,\ldots,0)
\ee
where $\dot{r}=\frac{dr}{dt}$. The proper velocity of the domain wall is
\be
 u^M=\frac{1}{\sqrt{A-B\dot{r}^2}}(1,\dot{r},0,\ldots,0).
\ee
The extrinsic curvature is $K_{MN}=h_M^P h_N^Q \nabla_P n_Q$ where
$h_{MN}=g_{MN}-n_Mn_N$.  We shall
compute it in the basis given by
$u_M,n_M,e^{(1)}_M,\ldots,e^{(D-2)}_M$ where the $e^{(i)}_M$ are an
orthonormal basis for the $D-2$ dimensional spatial sections.
In this basis, the $ij$ components of the extrinsic curvature are
\be
 K_{ij}=-\frac{\sqrt{AB}}{B}\frac{R'}{R}\frac{1}{\sqrt{A-B\dot{r}^2}}h_{ij},
\ee
where a prime denotes a derivative with respect to $r$. The $K_{00}$
component can be calculated as described in \cite{visser}.
Orthogonality of $u^M$ and $n_M$ gives
\be
 K_{00} = K_{MN}u^M u^N = u^M u^N \nabla _M n_N = -n_N u^M \nabla _M
 u_N = -n_M A^M,
\ee
where $A^M$ is the proper acceleration of the wall. Note $u_M A^M = 0$
so $A^M = \hat{A} n^M$ for
some $\hat{A}$. Hence $K_{00}=-\hat{A}$. To compute $\hat{A}$ we can
exploit the existence of a timelike Killing vector
$k=\frac{\partial}{\partial t}$. let $\tau$ denote proper time along
the wall. Then
\be
 \frac{d}{d\tau}(k^M u_M)=u^M u^N \nabla_M k_N+k_M A^M = k_M A^M = k^M
 n_M \hat{A},
\ee
where we have used Killing's equation.
Putting these results together gives
\be
\label{eqn:K00}
 K_{00} = \frac{1}{\dot{r}}{\sqrt{AB}}\frac{d}{dt}
 \left(\frac{A}{\sqrt{A-B\dot{r}^2}}\right).
\ee
(If $\dot{r}=0$ then $K_{00}=\frac{A'}{2A\sqrt{B}}$, which agrees with
the $\dot{r}\rightarrow 0$ limit of \ref{eqn:K00}.)

Having computed the extrinsic curvature we can now substitute into the
Israel equations. The $ij$ component gives
\be
\label{eqn:israelij}
 \frac{R'}{R} = \frac{\hat{V}(\phi)}{2(D-2)}\frac{\sqrt{AB}}{A}
 \sqrt{A-B\dot{r}^2},
\ee
which can now be used to eliminate $\sqrt{A-B\dot{r}^2}$ from the $00$
component. This yields
\be
 \left(\frac{R'}{R}\right)^{-1} \left(\frac{R'}{R}\right)' =
 \frac{(\hat{V}(\phi)\sqrt{AB})'}{\hat{V}(\phi)\sqrt{AB}}-\frac{R'}{R}.
\ee
This equation has to hold at every point visited by the domain
wall. Thus unless the domain wall remains at fixed $r$
(i.e. $\dot{r}\equiv 0$) then it has to hold over a range of $r$ and
can therefore be integrated, giving
\be
\label{eqn:ansatz}
 R'=C\hat{V}(\phi)\sqrt{AB},
\ee
where $C$ is a constant. 

We can now turn to the boundary condition on the dilaton. Equation
\ref{eqn:dilbc} can be simplified using equation \ref{eqn:israelij}
and reflection symmetry to give
\be
 \label{eqn:dilR}
 \frac{d\phi}{dR} = -\frac{D-2}{R} \frac{1}{\hat{V}}
 \frac{d\hat{V}}{d\phi},
\ee
which, if the wall visits a range of $R$, can be solved (in principle)
to yield $\phi$ as a function of $R$ without specifying the bulk potential.

Hence demanding that the domain wall be
non-static in a static bulk gives conditions relating the bulk metric
and dilaton. In the next section we shall use these conditions as
ans\"atze to solve the bulk Einstein equations.

\sect{The Bulk Metric}

\label{sec:bulk}

It is convenient to adopt the gauge $A(r)=B(r)^{-1}=U(r)$ for the
metric. The field equations are then
\be
\label{eqn:00plus11}
 \frac{R''}{R}=-\frac{1}{D-2}{\phi'}^2,
\ee
\be
\label{eqn:00}
 -\frac{D-2}{4}\frac{1}{R^{D-2}}\left(U'R^{D-2}\right)'=V,
\ee
\be
\label{eqn:ij}
 -\frac{1}{2R^{D-2}}\left(U\left(R^{D-2}\right)'\right)' +
\frac{k(D-2)(D-3)}{2R^2}=V,
\ee
\be
\label{eqn:phi}
 \frac{1}{R^{D-2}}\left(R^{D-2}U\phi'\right)'=\frac{dV}{d\phi}.
\ee
The ans\"atze \ref{eqn:ansatz} and \ref{eqn:dilR} can be employed to
seek solutions of these equations. They ensure that equation
\ref{eqn:00plus11} is satisfied. To proceed further it is necessary to
specify the domain wall potential.
We shall specialise to the case of a Liouville
potential:
\be
 \hat{V}(\phi)=\hat{V}_0 e^{\alpha\phi}.
\ee
Equations \ref{eqn:ansatz} and \ref{eqn:dilR} can be solved
simultaneously to give
\be
 \phi(r) = \phi_* -\frac{\alpha (D-2)}{\alpha^2(D-2)+1}\log r,
\ee
\be
 R(r) = (\alpha^2(D-2)+1)C\hat{V}_0 e^{\alpha\phi_*}
 r^{\frac{1}{\alpha^2(D-2)+1}},
\ee
where $\phi_*$ is a constant of integration. (A second constant of
integration can be set to zero by shifting the range of $r$.)

To make further progress it is necessary to specify the bulk
potential. We shall assume that this is also of Liouville type:
\be
 V(\phi)=V_0 e^{\beta\phi}.
\ee
Then equation \ref{eqn:ij} can be solved for $U(r)$. Substituting into
equations \ref{eqn:00} and \ref{eqn:phi}  yields constraints on the
parameters. There are three types of solutions. 

\bigskip

\begin{figure}
\begin{picture}(0,0)(0,0)
\put(60,70){\footnotesize\mbox{$V_0>0,M>0$}}
\put(135,70){\footnotesize\mbox{$V_0>0,M<0$}}
\put(210,70){\footnotesize\mbox{$V_0<0,M>0$}}
\put(285,70){\footnotesize\mbox{$V_0<0,M<0$}}
\end{picture}
\centerline{\psfig{file=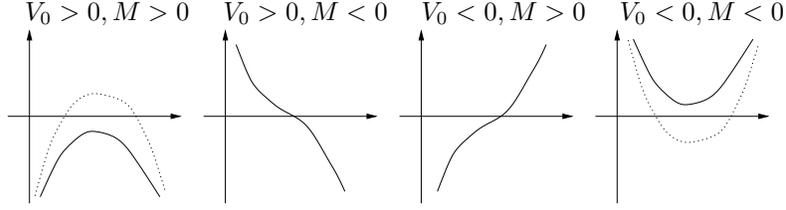,width=4.in}}
\caption{$U(R)$ for the Type I solutions. Dotted lines indicate
alternative behaviour.}
\label{fig:Imetric}
\end{figure}

Type I solutions
have $\alpha=\beta=0$, so the potential becomes a cosmological
constant. The solution has constant dilaton $\phi=\phi_0$. After
rescaling $t$ and changing variable from $r$ to $R$, the metric can be
written
\be
 ds^2=-U(R)dt^2+U(R)^{-1}dR^2+R^2d\Omega_k^2,
\ee
with
\be
 U(R)=k-2MR^{-(D-3)}-\frac{2V_0}{(D-1)(D-2)}R^2,
\ee
where $M$ is a constant. If $M=0$ then this is simply the metric of de
Sitter, Minkowski or anti-de Sitter spacetime according to the sign of
$V_0$. $U(R)$ is sketched in figure \ref{fig:Imetric} for $M\ne 0$.
It is easy to read off the horizon structure of the solutions from
these diagrams. 

When $V_0>0,M>0$ there are two possibilities. If $k=+1$ and
\be
\label{eqn:sdscond}
 \left[\frac{2V_0}{(D-2)(D-3)}\right]^{D-3} [(D-1)M]^2 < 1
\ee
(this corresponds to the dotted line in the first graph of figure
\ref{fig:Imetric})
then the solution is simply Schwarzschild-de Sitter, which has black
hole and cosmological horizons give by the two zeros of $U(R)$. If $k
\ne +1$ or equation \ref{eqn:sdscond} is not satisfied (corresponding
to the solid line) then the
solution is nowhere static ($R$ is a time coordinate) and there is a
cosmological singularity at $R=0$. At late times the metric approaches
that of de Sitter space.
 
If $V_0>0,M<0$ then there is a timelike naked singularity at
$R=0$. There is also a cosmological horizon (with geometry determined by
$k$) beyond which the metric is asymptotically de Sitter.

For $V_0<0,M>0$, there is a singularity at $R=0$ surrounded by an
event horizon beyond which the metric is asymptotically anti-de
Sitter. If $k=+1$ this is the Schwarzschild-anti de Sitter
solution. If $k=0$ or $k=-1$ then it describes a ``topological'' black
hole with a flat or hyperbolic event horizon. This can be made compact
by making identifications however this would also make the spatial
sections of the domain wall compact. 

When $V_0<0,M<0$, there are two possibilities. If $k=-1$ and
\be
\label{eqn:tadscond}
 \left[\frac{2|V_0|}{(D-2)(D-3)}\right]^{D-3} [(D-1)|M|]^2 < 1
\ee
(corresponding to the dotted line in the final graph of figure
\ref{fig:Imetric}) then the metric describes a topological black hole
in an asymptotically anti-de Sitter space, while if $k \ne -1$ or
equation \ref{eqn:tadscond} is not satisfied then it describes a timelike
naked singularity in an asymptotically anti-de Sitter space. 

\bigskip

\begin{figure}
\begin{picture}(0,0)(0,0)
\put(0,30){\footnotesize\mbox{$b^2>D-1$}}
\put(0,115){\footnotesize\mbox{$D-3<b^2$}}
\put(20,105){\footnotesize\mbox{$<D-1$}}
\put(0,195){\footnotesize\mbox{$b^2<D-3$}}
\put(60,240){\footnotesize\mbox{$V_0>0,M>0$}}
\put(135,240){\footnotesize\mbox{$V_0>0,M<0$}}
\put(210,240){\footnotesize\mbox{$V_0<0,M>0$}}
\put(285,240){\footnotesize\mbox{$V_0<0,M<0$}}
\end{picture}
\centerline{\psfig{file=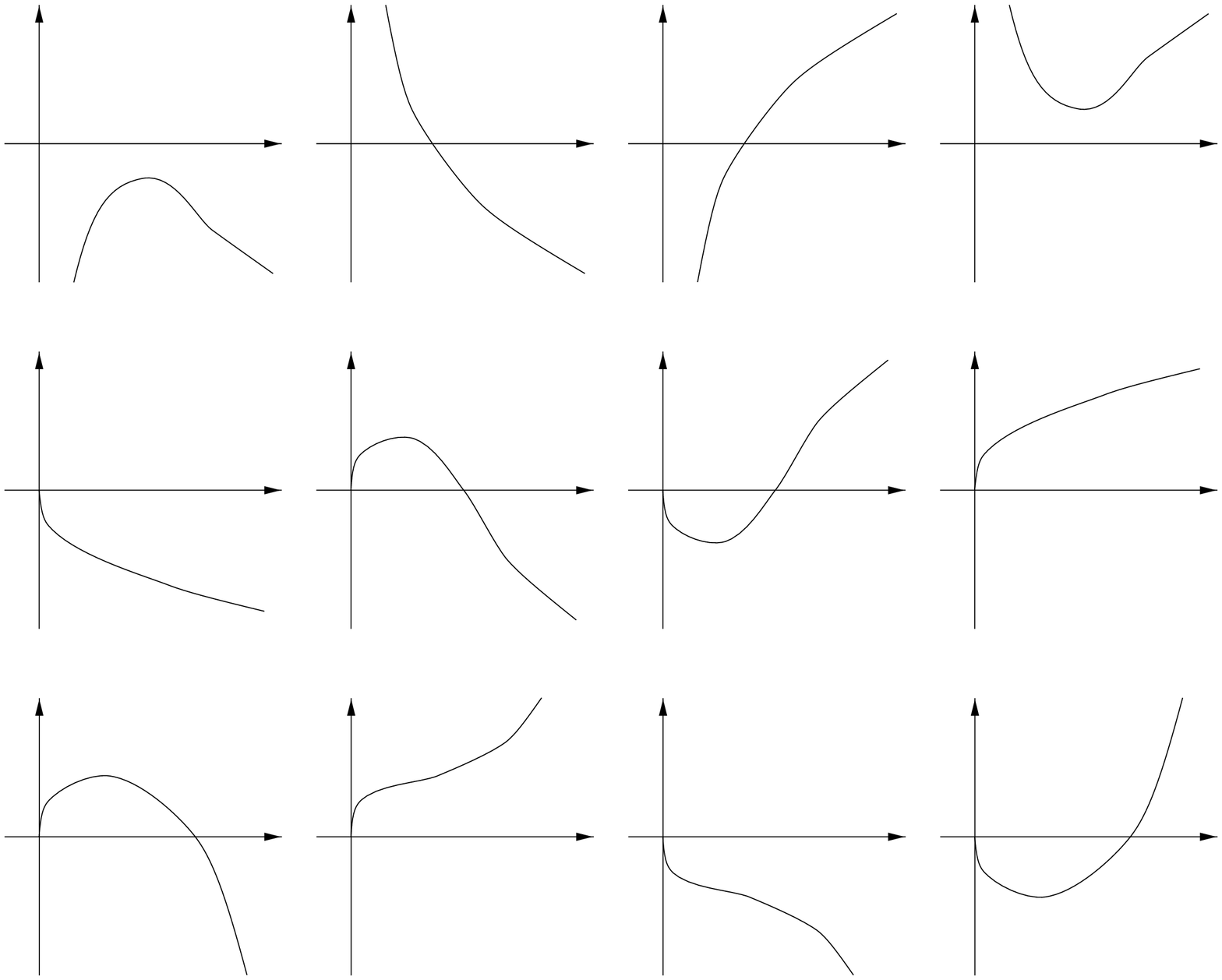,width=4.in}}
\caption{$U(r)$ for the Type II solutions.}
\label{fig:IImetric}
\end{figure}

Type II solutions have $\alpha=\beta /2$, $k=0$
and
\be
 U(r) = (1+b^2)^2 r^{\frac{2}{1+b^2}}
 \left(-2Mr^{-\frac{D-1-b^2}{1+b^2}} -\frac{2\Lambda}{(D-1-b^2)}\right),
\ee
\be
 R(r) = r^{\frac{1}{1+b^2}},
\ee
\be
 \phi(r)=\sqrt{D-2}\left(\phi_0-\frac{b}{1+b^2}\log r\right),
\ee
where  $M$ and $\phi_0$ are constants of integration and 
\be
 b=\frac{1}{2}\beta\sqrt{D-2},
\ee
\be
 \Lambda = \frac{V_0 e^{2b\phi_0}}{D-2}.
\ee
The function $U(r)$ is sketched in figure \ref{fig:IImetric}. 
These solutions (and their charged generalizations) were derived for
$D=4$ in \cite{cai} using the ansatz $R\propto r^N$. The solutions
with $M=0$ for arbitrary $D$ were derived in \cite{lu}.
Note that when $\alpha=\beta /2$, the theory becomes scale invariant
in the sense that a constant scale transformation $g_{MN} \rightarrow
\Omega^2 g_{MN}$, $\phi \rightarrow \phi-\frac{2}{\beta}\log\Omega$
simply multiplies the action by a constant: $S \rightarrow \Omega^{D-2}
S$. This means that the equations of motion are invariant under such a
transformation. 

All of the type II solutions are singular at $r=0$. For some the singularity is
timelike and for others it is spacelike. There is at most one horizon,
which is like a black hole horizon for some solutions and like a
cosmological horizon for others. The asymptotic (large $r$) behaviour
of the solutions depends on the value of $b^2$.

If $b^2<D-1$ then the asymptotic behaviour of the metric is determined
by the sign of $V_0$. If $V_0>0$ then $r$ is a time coordinate. By
rescaling $t$ and the flat spatial sections and changing variable
$r\rightarrow T(r)$, the metric can be written
as a FRW universe with flat spatial sections:
\be
\label{eqn:IIposV}
 ds^2\sim -dT^2+T^{\frac{2}{b^2}}d{\mathbf x}^2,
\ee
which is inflating if $b^2<1$.
If $V_0<0$ then $r$ is a spatial coordinate. By rescaling the other
coordinates and changing variable $r\rightarrow \rho(r)$, 
one can put the metric into the form
\be
\label{eqn:IInegV}
 ds^2\sim d\rho^2+\rho^{\frac{2}{b^2}}(-dt^2+d{\mathbf
x}^2),
\ee
which resembles the metric on anti-de Sitter space written in
horospherical coordinates. Note that when $M>0$ these solutions have 
a black hole type horizon and can be
interpreted as black $(D-2)$ brane solutions, generalizing some of the
supersymmetric solutions given in \cite{lu}. 

When $b^2>D-1$ the asymptotic behaviour of the metric is determined by
$M$. If $M>0$ then $r$ is a time coordinate and the metric has
anisotropic spatial sections and resembles a
Kasner solution
\be 
\label{eqn:IIposM}
 ds^2\sim -dT^2+T^{\frac{2(b^2-(D-3))}{b^2+D-1}}
dt^2 + T^{\frac{4}{b^2+D-1}}d{\mathbf x}^2,
\ee
where we have performed a transformation $r\rightarrow T(r)$ and
rescaled the other coordinates. The $t$ dimension expands faster than
the other three but none of the spatial dimensions inflate.
If $M<0$ then $r$ is a spatial coordinate and $t$ a time coordinate
and the metric is the same as above with the signs of the first two
terms changed. When $V_0<0$ (still with $M<0$), the metric 
has a black hole type horizon and hence can be interpreted as a topological
black hole with the curious property that the ``mass'' $M$ determines the
asymptotic structure.

\bigskip

\begin{figure}
\begin{picture}(0,0)(0,0)
\put(0,30){\footnotesize\mbox{$b^2>\frac{1}{D-3}$}}
\put(0,105){\footnotesize\mbox{$b^2<\frac{1}{D-3}$}}
\put(60,150){\footnotesize\mbox{$V_0>0,M>0$}}
\put(135,150){\footnotesize\mbox{$V_0>0,M<0$}}
\put(210,150){\footnotesize\mbox{$V_0<0,M>0$}}
\put(285,150){\footnotesize\mbox{$V_0<0,M<0$}}
\put(65,135){\footnotesize\mbox{$k=+1$}}
\put(140,135){\footnotesize\mbox{$k=+1$}}
\put(215,135){\footnotesize\mbox{$k=-1$}}
\put(290,135){\footnotesize\mbox{$k=-1$}}
\end{picture}
\centerline{\psfig{file=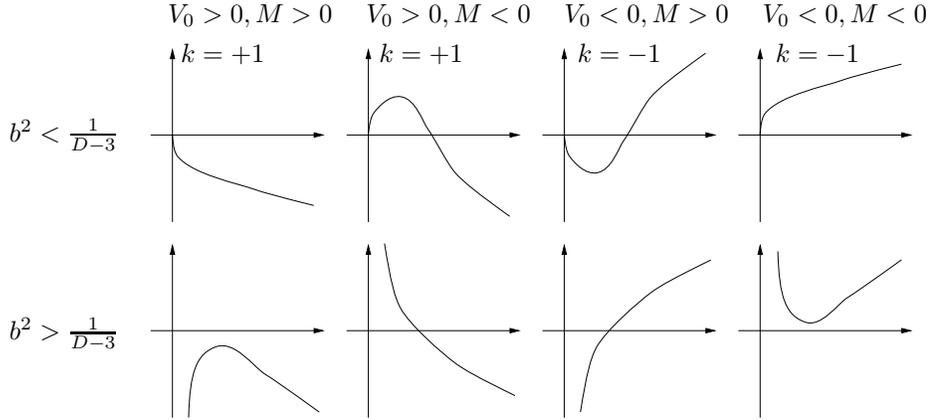,width=4.in}}
\caption{$U(r)$ for the Type III solutions. The value of $k$ in the
second row is the same as in the first row when $b^2<1$ and minus this
when $b^2>1$.}
\label{fig:IIImetric}
\end{figure}

Type III solutions have $\alpha=\frac{2}{\beta (D-2)}$. The metric is given by
\be 
 U(r)= (1+b^2)^2 r^{\frac{2}{1+b^2}}
 \left(-2Mr^{-\frac{1+b^2(D-3)}{1+b^2}} -\frac{2\Lambda}{(1+b^2(D-3))} \right),
\ee
\be
 R(r)=\gamma r^{\frac{b^2}{1+b^2}},
\ee
where
\be
 \gamma = \left(\frac{(D-3)}{2k\Lambda(1-b^2)}\right)^{\frac{1}{2}}
\ee
and $\phi(r)$ is the same as for the type II solutions. $b$ and
$\Lambda$ are defined as above and $k$ is given by the sign of
$\Lambda (1-b^2)$. (If $b^2=1$ then only the type II solution exists.) 
The function $U(r)$ is sketched in figure \ref{fig:IIImetric}.
These solutions (and their charged generalizations) were derived
in \cite{chan} for $k=+1$ and in \cite{cai} for $k=-1,D=4$ by making
the ansatz $r\propto r^N$.  
The solutions are all singular at $r=0$. Some have a horizon of
black hole or cosmological type. Their asymptotic (large $r$) behaviour
is determined by the sign of $V_0$. If $V_0>0$ then by rescaling $t$
and performing a change of variable $r\rightarrow T(r)$ the metric can
be written in an anisotropic cosmological form
\be
\label{eqn:IIIposV}
 ds^2 \sim
-dT^2+T^{\frac{2}{b^2}}dt^2+\frac{(D-3)b^4}{|1-b^2|(1+b^2(D-3))} T^2
d\Omega_k^2,
\ee
with $k$ given by the sign of $1-b^2$. When $b^2<1$ the spatial
sections have cylindrical topology and the axial ($t$) dimension
inflates but when $b^2>1$ the $t$ dimension grows more slowly 
than the other spatial dimensions.

If $V_0<0$ then $r$ is a spatial coordinate and the asymptotic solution
is the same as the above with the signs of the first two terms
changed i.e.
\be
\label{eqn:IIInegV}
 ds^2 \sim
 -\rho^{\frac{2}{b^2}}dt^2+d\rho^2+\frac{(D-3)b^4}{|1-b^2|(1+b^2(D-3))} \rho^2
d\Omega_k^2,
\ee
$k$ is given by the sign of $b^2-1$.

Note that only the $V_0<0,M>0$ solutions are of black hole type. Of
these, the $b^2<1$ solutions have hyperbolic spatial sections and the
$b^2>1$ solutions have round spatial sections. 

\bigskip

We have obtained solutions to the bulk field equations with
non-trivial dilaton by restricting
$\alpha$ to take one of two values determined by $\beta$. For each
value of $\beta$ there are two solutions, one with $k=0$ and the other
with $k=\pm 1$. We refer the reader to
references \cite{cai,chan} for analyses of the horizon structure and
thermodynamic properties of these solutions.

\sect{Analysis of Domain Wall Motion}

\label{sec:results}

The trajectory of the domain wall is determined by the Israel equation
\ref{eqn:israelij}. For the gauge used in the previous section, this
reduces to
\be
\label{eqn:israelij2}
 \frac{R'}{R}=\frac{\hat{V}(\phi)}
 {2(D-2)\sqrt{\left(\frac{dr}{d\tau}\right)^2+U}},
\ee
where $\tau$ is proper time on the domain wall world volume. 
There are two subtleties that we have to deal with
before studying solutions of this equation. The first is that the
solutions of the previous section all have $R'>0$, so the above
equation appears to rule out the possibility of $\hat{V}<0$. However
reversing the direction of the normal to the domain wall reverses the
sign on the right hand side. Equivalently, if $\hat{V}<0$ then the bulk
solutions derived above are valid on the $r>r(t)$ parts of the
surfaces of constant $t$ rather than the $r<r(t)$ parts. The geometrical
interpretation of this is that if one approaches a domain wall along
its normal then the spatial sections grow if the
wall has positive energy density but decrease if it has negative 
energy density.

The second subtlety arises when $U<0$. If horizons are present then
$U$ is positive in some (static) region but can be negative in other
regions. These can be dealt with in the standard way by introducing
Eddington-Finkelstein coordinates \cite{hawking}. The analysis of
section \ref{sec:israel} can be repeated in these coordinates,
reproducing equations \ref{eqn:israelij2} and \ref{eqn:ansatz},
which demonstrates their validity when the domain wall crosses a horizon.
If $U$ is negative everywhere then $r$ becomes the time coordinate and
it is convenient to reverse the direction of the normal so that the
bulk solution is valid on the $t<t(r)$ parts of surfaces of constant
$r$. Equations \ref{eqn:israelij2} and \ref{eqn:ansatz} then
become valid in this case too.

Equation \ref{eqn:israelij2} can be written
\be
\label{eqn:dweqn}
 \frac{1}{2}\left(\frac{dR}{d\tau}\right)^2+F(R)=0
\ee
where $\tau$ is proper time on the domain wall i.e. the induced metric
on the domain wall is
\be
 ds^2=-d\tau^2+R(\tau)^2 d\Omega_k^2.
\ee
This is the metric of a FRW universe. Equation \ref{eqn:dweqn}
determines the evolution of the scale factor $R(\tau)$ and is simply
the equation for a particle of unit mass and zero energy rolling in a 
potential $F(R)$. The potential $F(R)$ is given by
\be
 F(R)=\frac{1}{2}U{R'}^2-\frac{1}{8(D-2)^2}\hat{V}^2R^2.
\ee
Clearly solutions only exist when $F(R)\leq 0$. This is automatic if
$U<0$ i.e. if $r$ is a time coordinate. Inflationary solutions 
are of particular
interest. Inflation on the domain wall is defined by
$\frac{d^2R}{d\tau^2}>0$, which occurs when $\frac{dF}{dR}<0$.

In the next subsections we shall compute $F(R)$ for 
the solutions found in the previous section. 

\subsection{Type I Solutions}

These have
\be
 F(R)=\frac{k}{2}-MR^{-(D-3)}-\hat{\Lambda}R^2,
\ee
where the effective cosmological constant on the domain wall is
\be
 \hat{\Lambda} = \frac{1}{D-2} \left[\frac{V_0}{D-1} +
\frac{\hat{V}_0^2}{8(D-2)}\right].
\ee
There are several cases for which these solutions have been
extensively studied. Consider first the case $M=0$. T
he bulk spacetime is simply de
Sitter, Minkowski or anti-de Sitter space depending on the sign of
$V_0$. The position of the domain wall is given by solving equation
\ref{eqn:dweqn}. For $\hat{\Lambda}>0$, $R$
increases exponentially, corresponding to a de Sitter solution on the
domain wall world volume. When $\hat{V}_0>0$, the total bulk spacetime
is given by matching the bulk solution for $R<R(\tau)$ to a copy of 
itself across the
domain wall at $R=R(\tau)$. When $\hat{V}_0<0$, the $R>R(\tau)$ part
of the bulk
is matched to a copy of itself across the domain wall. An example is the
Vilenkin-Ipser-Sikivie domain wall \cite{ber,vil,ip}, 
which has Minkowski space in the bulk and spherical
spatial sections. For $\hat{V_0}>0$ the solution corresponds to gluing
the (flat) interior of a de Sitter hyperboloid embedded in Minkowski
space to a copy of itself while for $\hat{V_0}<0$ the exterior of the
hyperboloid must be used.

If $\hat{\Lambda}<0$ then a non-trivial solution
only exists for the case of open spatial sections. The domain wall can
expand to a maximum value of $R$ and then recollapse to $R=0$. The
world volume and bulk solutions are both anti-de Sitter.

When $\hat{\Lambda}=0$ the domain wall vacuum energy exactly cancels
the effect of the (negative) bulk cosmological constant 
(as far as the motion of
the wall is concerned) and the world volume metric is flat. Note that
$F \le 0$ requires $k=0$ or $k=-1$. In these cases the domain wall
is simply a horosphere of the bulk anti-de Sitter space. 

\bigskip

When $M \ne 0$
there is a singularity at $R=0$. If the domain wall has positive
energy density (i.e. $\hat{V}_0>0$) then the relevant part of the
bulk spacetime is $R<R(\tau)$, which contains the singularity. If it
has negative energy density then the relevant part is $R>R(\tau)$,
which is non-singular unless the wall reaches $R=0$. 

From any solution for the domain wall motion one can generate another
by time reversal. In what follows we shall only discuss one member of
each pair of time reversal related solutions. 

\begin{figure}
\begin{picture}(0,0)(0,0)
\put(60,70){\footnotesize\mbox{$\hat{\Lambda}>0,M>0$}}
\put(135,70){\footnotesize\mbox{$\hat{\Lambda}>0,M<0$}}
\put(210,70){\footnotesize\mbox{$\hat{\Lambda}<0,M>0$}}
\put(285,70){\footnotesize\mbox{$\hat{\Lambda}<0,M<0$}}
\end{picture}
\centerline{\psfig{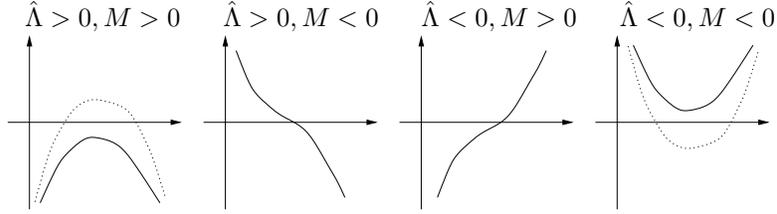}}
\caption{$F(R)$ for the Type I solutions. Dotted lines indicate
alternative behaviour.}
\label{fig:IF}
\end{figure}

$F(R)$ is plotted in figure \ref{fig:IF}. The qualitative behaviour of
the wall is easily read off from this figure. The four graphs
correspond to the following cases.

Case \emph{i}) $\hat{\Lambda}>0,M<0$. There are two possibilities. The
first (given by the dotted line in the first graph of figure \ref{fig:IF})
is $k=+1$ with
\be
 \label{eqn:fallin}
 (D-1) \left[\frac{2\hat{\Lambda}}{D-3}\right]^{(D-3)} M^2 < 1,
\ee
which corresponds to a Schwarzschild-de Sitter/anti-de Sitter bulk
solution. In this
case the wall can either expand out of the white hole region of the
bulk spacetime and recollapse into the black hole region, or it can
collapse from infinity, stop outside the black hole and re-expand to
infinity. 

If $k \ne +1$ or equation \ref{eqn:fallin} is not satisfied then the
domain wall emerges from a white hole or cosmological singularity
and expands to infinity.

Case \emph{ii}) $\hat{\Lambda}>0,M<0$. The bulk has a timelike naked
singularity. The domain wall collapses from infinity but is stopped by
the repulsive singularity. It then re-expands to infinity.

Case \emph{iii}) $\hat{\Lambda}<0,M>0$. The bulk solution is anti-de
Sitter space with a black hole or topological black hole. The domain
wall expands out of the white hole region and the negative
cosmological constant overwhelms the energy density of the domain
wall, causing it to recollapse into the black hole. 

Case \emph{iv}) $\hat{\Lambda}<0,M<0$, $k=-1$ and 
\be
 \label{eqn:repel}
 (D-1) \left[\frac{2|\hat{\Lambda}|}{D-3}\right]^{(D-1)} |M|^2 < 1,
\ee
The domain wall starts at a finite distance from a naked singularity,
which repels it, causing it to accelerate away. This expansion is
halted by the bulk (negative) cosmological constant and recollapse
occurs. The cycle then repeats. Thus the world volume of the domain
wall describes an open universe that undergoes a brief period of
inflation.

Note that the cases with $\hat{\Lambda}>0$ all expand to
infinity. This expansion is accelerating so the world volume undergoes
inflation. It is easy to see that the world volume solution approaches
de Sitter space at late times. 

The cases $k=0,-1$ with $V_0<0$ have been studied recently by Mann
\cite{mann}, who was interested in pair
creation of charged black holes with arbitrary event horizon topology.
His bulk configurations are slightly more general than our
Type I solutions because he allows for a single $U(1)$ charge in the bulk.
He studied the equations of motion for a domain wall in these backgrounds
because he was using the domain wall mechanism \cite{cham} to 
pair create the black holes. 

\begin{figure}
\begin{picture}(0,0)(0,0)
\put(0,30){\footnotesize\mbox{$b^2>D-1$}}
\put(0,115){\footnotesize\mbox{$1<b^2$}}
\put(15,105){\footnotesize\mbox{$<D-1$}}
\put(15,195){\footnotesize\mbox{$b^2<1$}}
\put(60,240){\footnotesize\mbox{$\hat{\Lambda}>0,M>0$}}
\put(135,240){\footnotesize\mbox{$\hat{\Lambda}>0,M<0$}}
\put(210,240){\footnotesize\mbox{$\hat{\Lambda}<0,M>0$}}
\put(285,240){\footnotesize\mbox{$\hat{\Lambda}<0,M<0$}}
\end{picture}
\centerline{\psfig{file=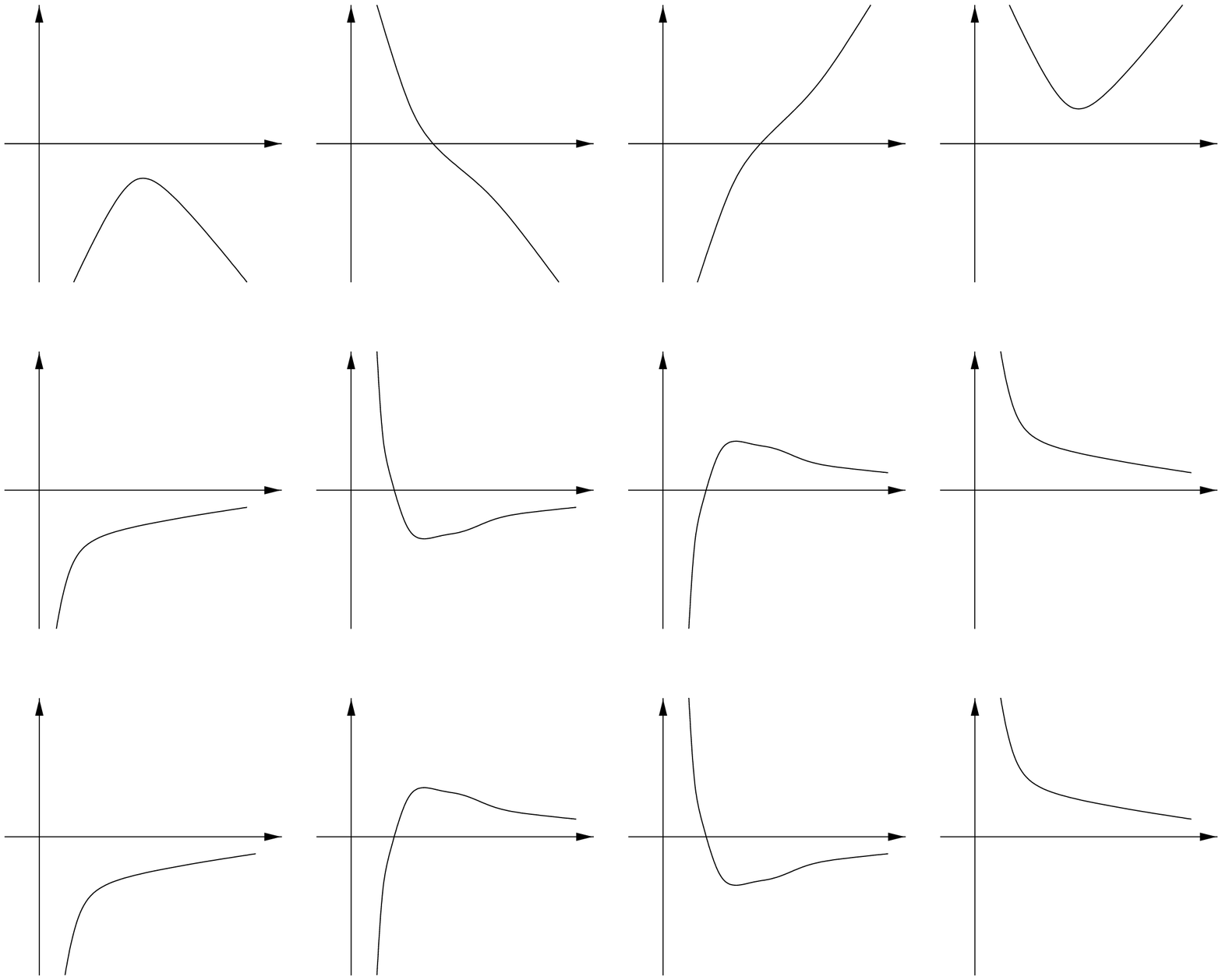,width=4.in}}
\caption{$F(R)$ for the Type II solutions.}
\label{fig:IIF}
\end{figure}

\subsection{Type II Solutions}

\be
 F(R)=-R^{2(1-b^2)} \left( MR^{-(D-1-b^2)} + \hat{\Lambda} \right),
\ee
where
\be 
 \hat{\Lambda} =  \frac{e^{2b\phi_0}}{D-2} \left( \frac{V_0}{D-1-b^2}
+ \frac{\hat{V}_0^2}{8(D-2)} \right).
\ee
$F(R)$ is sketched in figure \ref{fig:IIF}. Four classes of behaviour
 are apparent.

Class \emph{i}) $F(R)>0$ everywhere. No solutions exist.

Class \emph{ii}) $F(R)<0$ everywhere. These all have $\hat{\Lambda}>0$
and $M>0$ and describe a domain wall that emerges out of a singularity
at $R=0$ and expands forever. The singularity may be a timelike naked
singularity, a cosmological singularity or a white hole type
singularity hidden behind an event horizon. 

Class \emph{iii}) $F(R)$ positive for small $R$ and negative for large
$R$. These solutions describe a domain wall collapsing from infinity to a
minimum size (where $F(R)=0$) and then re-expanding to infinity. The
collapse is halted by a repulsive timelike naked singularity in the
bulk at $r=0$. 

Class \emph{iv}) $F(R)$ negative for small $R$ and positive for large
$R$. These solutions describe either a domain wall expanding out of a
timelike naked singularity at $R=0$ and then recollapsing into it, or
expanding out of a white hole type singularity and collapsing into a
black hole type singularity.

\bigskip

It is interesting to ask how the solutions that expand forever behave 
at late times. If $b^2<D-1$ then the domain wall can only expand
forever if $\hat{\Lambda}>0$. The scale factor at late times behaves as
$\tau^{\frac{1}{b^2}}$, which is inflationary if $b^2<1$. When $V_0>0$
this is the same behaviour as the scale factor of the $D-1$
dimensional spatial sections of the bulk (see equation
\ref{eqn:IIposV}) so the domain wall is simply comoving
with these spatial sections. If $V_0<0$ then the bulk metric is
static and the expansion is driven by the energy density of the domain
wall. In the coordinates of equation \ref{eqn:IInegV}, the position
of the domain wall is given by $\rho=\tau$ and $t=\frac{\sqrt{2}b^2}
{b^2-1} \tau^{1-\frac{1}{b^2}}$. When $b^2<1$, $t$ approaches a
constant at late proper time while if $b^2>1$ then $t$ becomes large
at late proper time.

When $b^2>D-1$, the large $R$ behaviour of both the domain wall and
the bulk is given by the sign of $M$. If $M<0$ then the domain wall
cannot expand indefinitely. If $M>0$ then at late times, $R$ is
proportional to $\tau^{\frac{2}{b^2+D-1}}$. Comparing this with the
behaviour of the bulk (equation \ref{eqn:IIposM}) we see that the
domain wall once again sits at a fixed position relative to the
expanding bulk spatial sections and the bulk expands fastest in the
direction transverse to the domain wall. 

\bigskip

There are two solutions which give a finite period of inflation. These
are the $\hat{\Lambda}>0,M<0,1<b^2<D-1$ solution and the
$\hat{\Lambda}<0,M>0,b^2>D-1$ solution. In both of these the domain
wall collapses from infinity, gets repelled by a timelike naked
singularity and then expands. Inflation occurs when the expansion
starts. In both cases the scale factor increases by a factor of
\be
 \left(\frac{D-3+b^2}{2(b^2-1)}\right)^{\frac{1}{D-1-b^2}},
\ee
which is cosmologically negligible unless $b^2$ is 
exponentially close to 1. 
Note that this expression is independent of $M$, a consequence 
of scale invariance.

\subsection{Type III Solutions}

\begin{figure}
\begin{picture}(0,0)(0,0)
\put(0,30){\footnotesize\mbox{$b^2>1$}}
\put(0,115){\footnotesize\mbox{$\frac{1}{D-1}<b^2$}}
\put(30,105){\footnotesize\mbox{$<1$}}
\put(5,195){\footnotesize\mbox{$b^2<\frac{1}{D-1}$}}
\put(60,240){\footnotesize\mbox{$V_0>0,M>0$}}
\put(135,240){\footnotesize\mbox{$V_0>0,M<0$}}
\put(210,240){\footnotesize\mbox{$V_0<0,M>0$}}
\put(285,240){\footnotesize\mbox{$V_0<0,M<0$}}
\put(75,55){\footnotesize\mbox{$k=-1$}}
\put(150,55){\footnotesize\mbox{$k=-1$}}
\put(225,55){\footnotesize\mbox{$k=+1$}}
\put(300,55){\footnotesize\mbox{$k=+1$}}
\put(75,135){\footnotesize\mbox{$k=+1$}}
\put(150,135){\footnotesize\mbox{$k=+1$}}
\put(225,135){\footnotesize\mbox{$k=-1$}}
\put(300,135){\footnotesize\mbox{$k=-1$}}
\put(75,220){\footnotesize\mbox{$k=+1$}}
\put(150,220){\footnotesize\mbox{$k=+1$}}
\put(225,220){\footnotesize\mbox{$k=-1$}}
\put(300,220){\footnotesize\mbox{$k=-1$}}
\end{picture}
\centerline{\psfig{file=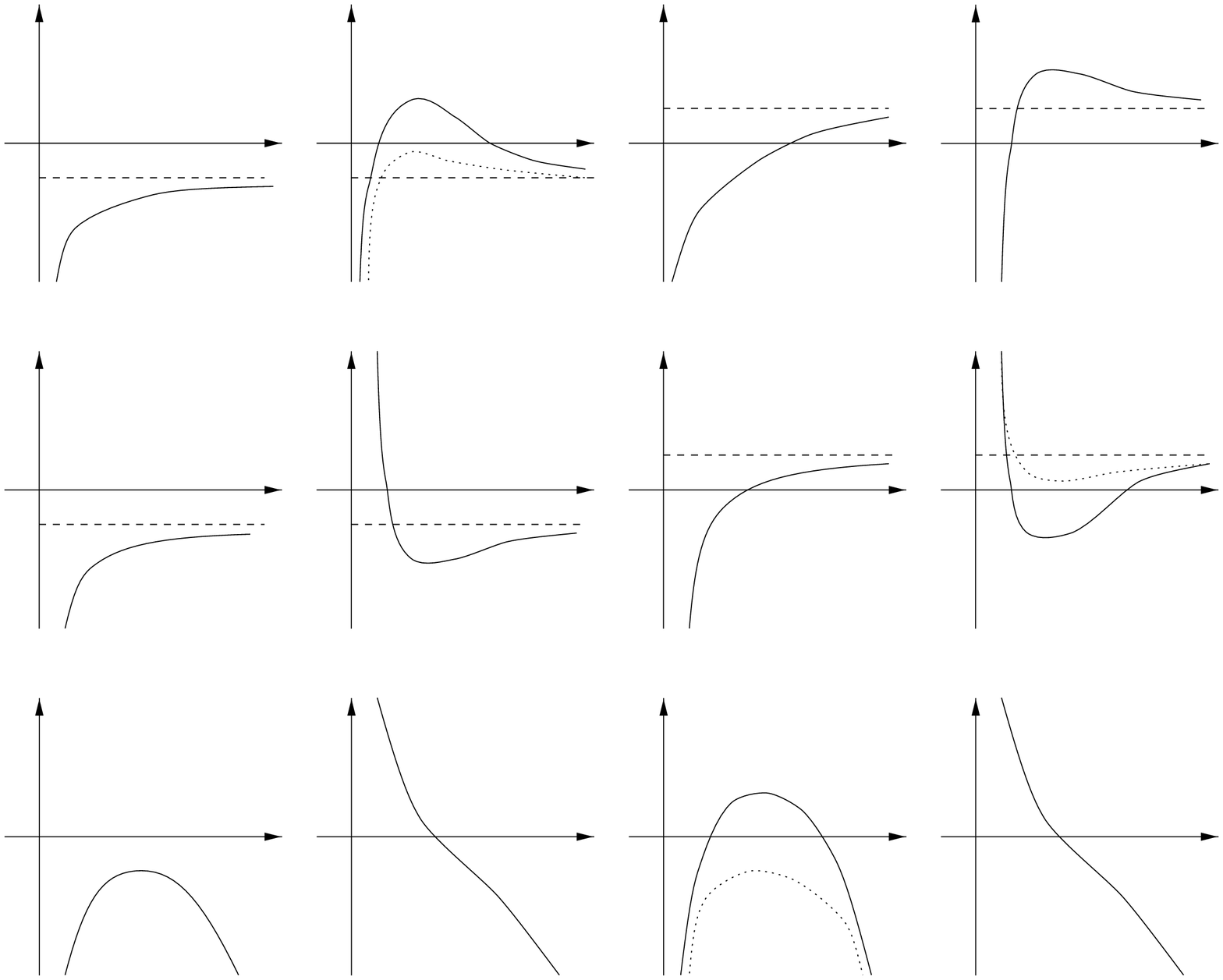,width=4.in}}
\caption{$F(R)$ for the Type III solutions. Dashed lines indicate
asymptotes. Dotted lines indicate alternative behaviour.}
\label{fig:IIIF}
\end{figure}

For the type III solutions,
\ba
\label{eqn:IIIF}
 F(R) = &-&\frac{(D-3) b^4}{2k(1-b^2)(1+b^2(D-3))} - {} \nonumber \\ {}
&-&  M\gamma^2 b^4
\left(\frac{R}{\gamma}\right)^{-\left(D-3+\frac{1}{b^2}\right)} -
\frac{\hat{V_0}^2 e^{\frac{2\phi_0}{b}}\gamma^2}{8(D-2)^2}
\left(\frac{R}{\gamma}\right)^{-2\left(\frac{1}{b^2}-1\right)}.
\ea
This is sketched in figure \ref{fig:IIIF}.
The behaviour can be divided into five classes.

Class \emph{i}) $F(R)$ is negative everywhere. In these solutions the
domain wall expands out of a cosmological singularity, timelike naked
singularity or white hole singularity and expands forever. 

Class \emph{ii}) $F(R)$ is negative for small $R$ and positive for
large $R$. The domain wall either expands out of a timelike naked
singularity and recollapses into the same singularity, or it expands
out of a white hole singularity and collapses into a black hole
singularity.

Class \emph{iii}) $F(R)$ is positive for small $R$ and negative for
large $R$. The domain wall collapses from infinity to a finite size
and then re-expands to infinity. In most cases the collapse is halted
by a timelike naked singularity. The exception is if $b^2>1,V_0<0$ and
$M$ is positive but less than a calculable upper bound (corresponding
to the solid line in the relevant graph of figure \ref{fig:IIIF}). Then the
turning point occurs outside the horizon of a spherical black hole.

Class \emph{iv}) $F(R)$ is positive for a finite range of $R$. This
only occurs when 
$\frac{1}{D-1}<b^2<1$, $V_0<0$ and $M$ is negative
but greater than a calculable lower bound (corresponding to the solid
line in the relevant graph of figure \ref{fig:IIIF}). 
The domain wall is repelled by a naked timelike singularity,
inflates for a brief period, decelerates to a halt, recollapses and
then repeats this cycle. The world volume describes an open
`bouncing' universe.

\bigskip

The solutions which expand to infinity exhibit late time behaviour of
two different types. When $b^2<1$ and $V_0>0$, the first term in $F$
is dominant at late times. This term can be thought of as the term
arising from the bulk curvature. All of these solutions have spherical
spatial sections and at late times the scale factor grows linearly
with proper time
\be 
 R(\tau) \sim \left[ \frac{(D-3)b^4}{(1-b^2)(1+b^2(D-3))}
\right]^{\frac{1}{2}} \tau.
\ee
If we compare this with the asymptotic behaviour of the bulk metric,
given by equation \ref{eqn:IIIposV}, it is clear that the domain wall
is simply comoving with the bulk i.e. it lies at fixed $t$. Recall
that the topology of the bulk spatial sections is cylindrical. The
domain wall remains at a fixed position on the axis of this cylinder
as it expands. Note that inflation occurs in the bulk in the direction
transverse to the domain wall. 

The second type of behaviour occurs for $b^2>1$, when the domain wall energy
density becomes dominant at late times. These solutions undergo power
law inflation on the world volume with the scale factor growing
proportionally to $\tau^{b^2}$.
When $V_0>0$, the coordinates of the
domain wall with respect to the bulk spacetime (give by equation
\ref{eqn:IIIposV}) are $T\propto\tau^{b^2}$ and
$t\propto\tau^{b^2-1}$, so the wall moves relative to the bulk spatial
sections. When $V_0<0$, the bulk spacetime is given by equation
\ref{eqn:IIInegV} and the position of the domain wall by
$\rho\propto\tau^{b^2}$ and $t\propto\tau^{b^2-1}$. 

\bigskip

It is also possible to have world volume inflation when $b^2<1$. Both of the
solutions with $\frac{1}{D-1}<b^2<1$ and $M<0$ have finite period of
inflation caused by the repulsive effect of a timelike
singularity. In the $V_0<0$ case, the amount of inflation is
negligible. In the $V_0>0$ case, a significant amount of inflation
only occurs if the dimensionless quantity
\be
 \frac{|M|V(\phi_0)}{\hat{V}(\phi_0)^2}
\ee
is exponentially large, corresponding to a very strong
singularity or very small boundary potential.

If $b^2<\frac{1}{D-1}$, $V_0>0$ and $M<0$ then an infinite amount of
inflation is possible. However this inflation dies out, with the
expansion approaching constant velocity at late times. A significant
period of rapid inflation would require exponential tuning as above. 

\sect{Dimensional Reduction}

\label{sec:higherdim}

Liouville potentials typically arise from dimensional reduction. To
see which couplings can arise in this manner, consider a model in
$D+n$ dimensions with action
\be
 S=\frac{1}{8\pi\bar{G}} \int d^{D+n}\bar{x} \sqrt{-\bar{g}}
(\frac{1}{2}\bar{R} - V_0) - \frac{1}{8\pi\bar{G}}\int_{\Sigma}
d^{D+n-1}\bar{x} \sqrt{-\bar{h}} (\{\bar{K}\} + \hat{V}_0),
\ee
where bars denote $(D+n)$-dimensional
quantities. The domain wall has a simple Nambu-Goto action with
tension $\hat{V}_0$. 
We can dimensionally reduce this using the ansatz
\be
 ds^2=e^{2A(x)}g_{MN}(x)dx^M dx^N + e^{2B(x)}g_{mn}(y)dy^m dy^m,
\ee
where $g_{MN}$ and $G_{mn}$ are the metrics on $D$-dimensional
spacetime and a $n$-dimensional internal Einstein space respectively and 
$B(x)=-\frac{(D-2)}{n}A(x)$ (to obtain the reduced action in the Einstein
frame).
The bulk action reduces to
\be
 S_{\mathrm{bulk}} = \int d^D x \sqrt{-g} \left(\frac{1}{2}R -
\frac{1}{2}(\partial \phi)^2 - \frac{\beta_1}{2}\nabla^2\phi - V_0
e^{\beta_1\phi} + \frac{1}{2} R_n e^{\beta_2\phi} \right),
\ee
where $R_n$ is the Ricci scalar of the
$n$-dimensional space,
\be
 \beta_1=2\left(\frac{n}{(D-2)(D+n-2)}\right)^{\frac{1}{2}}, \qquad
\beta_2 = 2 \left(\frac{(D+n-2)}{n(D-2)}\right)^{\frac{1}{2}} 
\ee
and $\phi(x)=2A(x)/\beta_1$. We have chosen units
so that $8\pi G = 1$, where $G$ is the $D$ dimensional Newton constant.
Note that if we define the parameters $b_i=\frac{1}{2}\beta_i\sqrt{D-2}$
as in section \ref{sec:bulk} then we have $b_1 = 1/b_2$. 

If the normal to the domain wall points only in directions
corresponding to the $D$-dimensional space then the domain wall action
reduces to
\be
 S_{DW} = -\int d^{D-1} x\sqrt{-h}\left( \left\{K +
\frac{\beta_1}{2}n.\partial\phi\right\} + \hat{V}_0 e^{\alpha
\phi} \right),
\ee
where $\alpha= \frac{\beta_1}{2} = \frac{2}{\beta_2(D-2)}$. The total
derivative in the bulk action cancels the normal derivative in the
domain wall action (recall that $n$ points \emph{into} the bulk). 
The bulk action has a sum of Liouville
potentials. The boundary action has a single Liouville potential. If
either $R_n=0$ or $V_0=0$ then the model reduces to the one we have
considered in previous sections and admits solutions of type II or
type III respectively. 

This implies that some of our type II (with $b^2<1$) and type III
solutions (with $b^2>1$) can be
oxidized to higher dimensional models in which the domain wall and
bulk actions are much simpler. Oxidation of a type II solution yields
a type I solution in $(D+n)$ dimensions with a non-vanishing bulk
cosmological constant and spatial sections that are products of
$D$-dimensional flat space with a $n$-dimensional Ricci flat space. 
Oxidation of a type III solution yields a type I solution with
vanishing bulk cosmological constant and spatial sections that are
products of a $D$-dimensional sphere (hyperboloid) 
with a positively (negatively) curved $n$-dimensional Einstein space.

In terms of the higher dimensional theory, the components of the
Einstein equations on the internal space give rise to the scalar
equation of motion of the reduced theory. The components of the Israel
conditions on the internal space give equation \ref{eqn:dilbc},
describing the jump in the scalar field at the domain wall.

Note that the scale transformations of the type II solutions obtained
by dimensional reduction leave the $(D+n)$ dimensional metric
invariant provided one assumes that the metric on the (Ricci flat) 
internal space scales in a suitable way. This symmetry arises as a
result of the dimensional reduction ansatz. (The cosmological
constant and domain wall tension break the scale invariance of the
higher dimensional action.)

The results of this section suggest that it would be possible to
generalize our method to deal with a bulk potential consisting of a
sum of Liouville potentials with parameters
$b=\frac{1}{\alpha\sqrt{D-2}}$ and $1/b$. In fact it
is straightforward to show that for a Liouville potential on the wall,
this is the most general bulk potential for which our method will
work\footnote{Actually there is a special limiting case, namely
$\alpha^2=1/(D-2)$ (i.e. $b^2=1$), 
for which the bulk potential takes the form $(V_0
+ V_1\phi) e^{\pm\frac{2\phi}{\sqrt{D-2}}}$.}. 
The bulk solutions in this case all have $k=\pm 1$. The $k=+1$
solutions were constructed in \cite{chan}. 

\sect{Ho\v{r}ava-Witten Cosmology}

\label{sec:hw}

An interesting scenario for which our method can be used to find
solutions is that of strongly coupled $E_8 \times E_8$ heterotic
string theory, which has been identified by Ho\v{r}ava and Witten
with M-theory compactified on a $S^1/Z_2$ orbifold with $E_8$ gauge
fields living on each orbifold fixed plane \cite{hw1,hw2}. 
This can be compactified to five dimensions on a Calabi-Yau space
\cite{witten}. Matching the predicted values for the four dimensional
gravitational and GUT couplings leads one to the conclusion that the
orbifold is an order of magnitude larger than the Calabi-Yau space
\cite{witten,banks}. Our universe is identified with one of the
orbifold fixed planes. The other fixed plane describes a ``shadow''
universe that interacts with our own only via bulk fields. 

Lukas \emph{et al} have shown that this five dimensional theory admits
a supersymmetric solution describing a pair of domain walls (the
orbifold fixed planes) \cite{lukas1}. 
Simple cosmological solutions have also been found by
separating variables in the bulk spacetime \cite{lukas2,hsr}. We can
apply the methods of the previous sections to find further solutions.

The model of Lukas \emph{et al} has $\alpha = \frac{\beta}{2} =
-\sqrt{2}$, $V_0 = \frac{a^2}{6}$ and $\hat{V}_0=\mp\sqrt{2}a$, where
$a$ is a constant related to the number of units of four-form flux on
the internal Calabi-Yau space, and the two sign choices refer to
domain walls of negative and positive tension, which we shall
call $M_1$ and $M_2$ respectively. Note that this theory cannot be
obtained by dimensional reduction of a theory with a cosmological constant.
The theory is scale invariant, so a type II ($k=0$)
solution exists. The bulk solution is described by
\be
 U(r) = 49r^{\frac{2}{7}}\left( \frac{a^2}{18}e^{-\sqrt{6}\phi_0} -
2Mr^{\frac{2}{7}} \right),
\ee
\be
 R(r)=r^{\frac{1}{7}},
\ee
\be
 \phi(r) = \sqrt{3}\left(\phi_0 + \frac{\sqrt{6}}{7}\log r\right).
\ee
The effective potential for the domain walls simplifies considerably:
\be
 F(R)=-MR^{-8},
\ee
from which we see that $M\ge 0$ is necessary for domain wall solutions
to exist in this bulk. If $M=0$ then $F\equiv 0$ and the domain walls
can be put anywhere and will remain static. The bulk solution has a
timelike naked singularity at $r=0$. The spatial sections must
decrease towards $M_1$ and
increase towards $M_2$ so if the former is at $r=r_1>0$ then
we must put the latter at $r=r_2>r_1$. The
bulk spacetime that is left after imposing reflection symmetry in each
domain wall is $r_1<r<r_2$, which is non-singular. This solution is
simply the supersymmetric domain wall solution of Lukas \emph{et al}.

When $M>0$ there is a timelike naked singularity at $R=0$ and a cosmological
horizon at $R = \frac{1}{6} |a| \sqrt{M}
e^{\frac{\sqrt{6}\phi_0}{2}}$. Static domain wall solutions are no
longer possible. The position of $M_i$ is given by
\be
 R_i(\tau) = \left(5\sqrt{2M}(\tau_i \pm \tau)\right)^{\frac{1}{5}},
\ee
where the $\tau_i$ are constants. The scale
factors on each domain wall grow very slowly with the velocity
approaching zero at late times. There are two choices of sign for each
domain wall, leading to a total of four solutions. In an obvious
notation these can be classified as followed:

$(++)$ solution. The $+$ sign occurs for both walls so they are both
moving outwards. We need $\tau_1<\tau_2$ for $M_1$ to occur at the 
smaller value of $R$. This
domain wall becomes singular at a finite proper time in its past. At
late times the separation between the two domain walls tends to zero
as $\tau^{-\frac{4}{5}}$. The $(--)$ solutions is simply the time
reversed version of this.

$(-+)$ solution. $M_1$ has $R$
decreasing and $M_2$ has $R$ increasing. They must have been
coincident at some time in the past then
moved apart with one falling into the singularity. The $(+-)$ solution
is the time reverse of this, which has $M_{1}$ emerging from the
singularity then colliding with $M_{2}$ which is moving towards the
singularity.

Since the distance between the universe and the shadow universe 
gives the string coupling, $(-+)$ solutions describe the
``pair creation'' of a universe-shadow universe pair from a region 
of very weak string coupling.  Likewise, the $(+-)$ solutions
describe the annihilation of such a pair.  It is unclear how we can
assign a rate, or probability, for such processes.

\sect{Conclusions}

\label{sec:conclude}

If a domain wall couples to a bulk matter field then one would expect
the bulk spacetime to be time dependent. 
We have investigated the conditions under which it is
possible to have such a domain wall moving in a \emph{static} bulk
spacetime with a dilaton. For the case of Liouville potentials in the
bulk and on the wall we have found that if the bulk and boundary
exponents are related in a certain ways then solutions can be found. 

The bulk solutions we have found are all singular and have at most one
horizon of black hole or cosmological type. Moving across such a
horizon takes one to a non-static region. This allowed us to study
dilatonic domain walls moving in a time dependent bulk. The behaviour
of the domain walls is qualitatively similar to the constant dilaton
solutions (type I) in certain respects i.e. they can fall into the 
singularity or expand forever. However in other respects our solutions
are different. For example, all of the type I solutions that expand to
infinity undergo inflation whereas we have found new solutions that
decelerate to a constant velocity at infinity.

\bigskip

There has been recent interest in cosmological models in which our
universe is viewed as a brane moving in a higher dimensional spacetime
\cite{large1}. 
In light of this, we have concentrated on the issue
of inflation on the domain wall world volume. If the domain wall
inflates then it must either move in the bulk or the bulk must be
inflating too. Most recent work has concentrated on the latter
possibility. We have solutions describing both cases. We believe that our
model is of interest because the inflaton couples to the domain wall
but is not restricted to its world volume. For the type II solutions,
power law inflation occurs when $b^2<1$ and for the type III solutions
when $b^2>1$. In the former case the bulk is also inflating (if
$V_0>0$) while in the latter it is not. The critical value for
inflation in both cases can be expressed as $\alpha^2<\frac{1}{D-2}$,
so it appears that it is the domain wall coupling rather than the bulk
coupling that dictates whether inflation occurs. (Note that the
critical value for inflation in the bulk is $\beta^2<\frac{4}{D-2}$.)
These inflationary solutions are similar to the Vilenkin-Ipser-Sikivie
domain wall \cite{ber,vil,ip} (which undergoes exponential inflation) in
the respect that inflation occurs because the domain wall energy
dominates bulk effects. However, in our solutions the energy density of
the domain wall tends to zero at late times but inflation continues. 
(A similar effect occurs in bulk inflation from an exponential
potential.)

\bigskip

We have shown that dimensional reduction of a theory consisting of
Einstein gravity with a cosmological constant and domain walls with
Nambu-Goto type actions gives a theory of the type we have considered
in this paper. Interestingly the dimensionally reduced theories always
have $\alpha^2<\frac{1}{D-2}$, so domain wall inflation arises
naturally from dimensional reduction. This is presumably because the
domain wall spacetime in the higher dimensional theory is
inflating. However exponential inflation in the higher dimensional
theory becomes power law inflation in the reduced theory.  

Our method can be used to find new cosmological solutions of
Ho\v{r}ava-Witten theory. These solutions do not appear very
phenomenologically interesting. In order to obtain inflation in this
model it seems necessary 
to include matter restricted to the world volume of the domain walls
\cite{lukas3}.

\bigskip

We would like to finish by mentioning possible generalizations and
applications of our work. 
It would be interesting to see if our method could be extended to
find solutions with different potentials in the bulk or on the
boundary. An obvious generalization would be to investigate further
the domain
wall motion when the bulk dilaton potential is a sum of Liouville
potentials with parameters $b$ and $1/b$, as obtained from dimensional
reduction of type I solutions. 
Other possible generalizations could include putting charge in the
bulk (such solutions were described in \cite{cai,chan}) or including
matter fields restricted to the domain wall world volume.  

Another obvious way of generalizing our work would be to relax the
assumption of reflection symmetry in the domain wall. One could also
allow the bulk potential to vary discontinuously across the wall, as
studied for the case of a cosmological constant in
\cite{ber,guth}. Such a scenario arises in type IIA string theory when
D8 branes are considered. Romans \cite{romans} found a ``massive''
generalization of the type IIA supergravity theory. His theory
contains a Liouville type potential with an arbitrary (positive)
coefficient. This was reformulated in \cite{berg} in terms of a ten
form field strength of a nine form potential that the D8 branes of
type IIA string theory couple to. The expectation value of this ten
form jumps across a D8-brane. This can be viewed as a jump in the
coefficient of a bulk Liouville potential. Our method might be useful
in obtaining solutions describing dynamic D8 branes in this theory.

The main question concerning our solutions is whether they
are stable. In other words, if the domain wall is perturbed then will the
perturbation remain small or will it grow and change the bulk metric,
for example converting a static metric to a time-dependent one? If
they are stable then it would be interesting to examine how 
perturbations behave in this model, especially as viewed by an
inhabitant of the domain wall world volume. 

\bigskip

\centerline {\bf Acknowledgements}

We have enjoyed useful conversations with Gary Gibbons and Stephen Hawking.
HAC was supported by Pembroke College, Cambridge.

\end{document}